\documentclass[twocolumn]{aastex63}
\usepackage{lineno}

\shorttitle{Paper Draft}
\shortauthors{Hatcher et al.}

\graphicspath{{./}{figures/}}
\usepackage{xcolor}

\begin{document}

\title{Where Do Obscured AGN Fit in a Galaxy's Timeline?}

\author{Cassandra Hatcher}
\affiliation{Department of Physics and Astronomy, the University of Kansas, Lawrence, KS. 66045, USA}
\email{hatcher.c@ku.edu}

\author{Allison Kirkpatrick}
\affiliation{Department of Physics and Astronomy, the University of Kansas, Lawrence, KS. 66045, USA}

\author{Francesca Fornasini}
\affiliation{Physics and Astronomy Department, Stonehill College, 320 Washington St., North Easton, MA 02357}
\affiliation{Center for Astrophysics, Harvard \& Smithsonian, 60 Garden Street, Cambridge, MA 02138}

\author{Francesca Civano}
\affiliation{Center for Astrophysics, Harvard \& Smithsonian, 60 Garden Street, Cambridge, MA 02138}

\author{Erini Lambrides}
\affiliation{Department of Physics and Astronomy, Johns Hopkins University, Baltimore, MD 21218, USA}

\author{Dale Kocesvski}
\affiliation{Department of Physics and Astronomy, Colby College, Waterville, ME 04961, USA}

\author{Christopher M. Carroll}
\affiliation{Department of Physics and Astronomy, Dartmouth College, 6127 Wilder Laboratory, Hanover, NH 03755, USA}

\author{Mauro Giavalisco}
\affiliation{Department of Astronomy, University of Massachusettes, Amherst, MA 01003, USA}

\author{Ryan Hickox}
\affiliation{Department of Physics and Astronomy, Dartmouth College, 6127 Wilder Laboratory, Hanover, NH 03755, USA}

\author{Zhiyuan Ji}
\affiliation{Department of Astronomy, University of Massachusettes, Amherst, MA 01003, USA}

\begin{abstract}

Many X-ray bright active galactic nuclei (AGN) are predicted to follow an extended stage of obscured black hole growth. In support of this picture we examine the X-ray undetected AGNs in the COSMOS field and compare their host galaxies with X-ray bright AGNs. We examine galaxies with $M_\ast>10^{9.5}M_\odot$ for the presence of AGNs at redshifts $z=0.5-3$. We select AGNs in the infrared using \textit{Spitzer} and \textit{Herschel} detections and use color selection techniques to select AGNs within strongly star forming hosts. We stack \textit{Chandra} X-ray data of galaxies with an IR detection but lacking an X-ray detection to obtain soft and hard fluxes, allowing us to measure the energetics of these AGNs. We find a clear correlation between X-ray luminosity and IR AGN luminosity in the stacked galaxies. We also find that X-ray undetected AGNs all lie on the main sequence -- the tight correlation between SFR and $M_\ast$ that holds for the majority of galaxies, regardless of mass or redshift. This work demonstrates that there is a higher population of obscured AGNs than previously thought. 

\end{abstract}


\section{Introduction \label{sec:intro}}

There is a strong relationship between a host galaxy’s stellar content and the mass of its central supermassive black hole (SMBH), although the physical mechanisms causing this relationship are not yet understood \citep{Silk1998, Ferrarese2000, Gerbhardt2000, Treister2012, Kormendy2013}. A SMBH that is actively accreting material from its host galaxy is categorized as an active galactic nucleus (AGN). Essentially all massive galaxies host a SMBH, but not all are visible as AGN. Star forming galaxies (SFGs) lack a discernable AGN, and while they are presumably still hosting a central SMBH, star formation is the main source of energy being emitted. It is still unclear how or when in a galaxy’s evolution these AGNs become activated \citep{King2003, Sun2015}. Some AGNs can be triggered during a merger, but for non-merging galaxies, AGN activation is still an open question \citep{Treister2012}, though there is supporting evidence that internal secular evolution is the predominate mechanism for triggering AGN activity \citep{Chang2017, Man2019}.

A very powerful AGN may quench star formation in its host galaxy through energetic winds or radiation \citep{Treister2012}. There is growing evidence to support the association between AGNs and quenching, as many AGNs are 
found in the transition area between star-forming and passive populations \citep{Wang2017, Gu2018, Man2019}. Some of the most luminous AGNs are found in galaxies with highly depleted star formation rates (SFRs), indicating quenching, but what is still unknown is the average host galaxy properties of lower-luminosity AGN ($L_{X} < 10^{42}$ ${\rm erg}$ ${\rm s}^{-1}$). However, if we can measure the accretion rate onto all SMBHs, with both high and low luminosities, then we can determine if there is a universal relationship between the growth of SMBHs and SFRs \citep{Volonteri2015}. 

Evolved galaxies formed the majority of their black hole and stellar mass at $1 < z < 3$, making these sources imperative to our understanding of the interactions between AGN growth and its host \citep{Madau2014}. Current studies at these redshifts are of luminous AGN ($L_{X} > 10^{42}$ ${\rm erg}$ ${\rm s}^{-1}$); there is limited data on lower-luminosity AGNs due to the long integration times needed for detection. \cite{Dekel2009} found evidence that X-ray luminous AGNs (X-ray AGNs) occur after a galaxy has built up its central stellar mass density, a processes called compaction, but prior to star formation quenching. According to this framework, X-ray AGNs occur in a special phase of a galaxy's life. Popular theories of quasar activation, based on gas-rich galaxy mergers, suggest there should be an extended stage of obscured black hole growth \citep{Hopkins2006}. A large fraction of AGNs are predicted to grow behind obscuring dust and gas in population synthesis models \citep{Gilli2007}. \cite{Kocevski2015} found support that obscured AGNs are a distinct phase of SMBH growth that follows a merger/interaction event. It is during this obscured phase that AGNs are predicted to accrete the bulk of their mass and produce most of their feedback into their host galaxies \citep{Hopkins2006}. It is predicted that this AGN feedback can result in the expulsion of gas and/or prevent the gas from cooling and collapsing into stars \citep{Bower2006, Croton2006}. 

AGNs emission can be detected in a large range of wavelengths, from radio to gamma-rays. Method efficiency is highly dependent on AGN obscuration \citep{Hickox2018}. Obscured AGNs can be missed by many surveys, such as rest-frame UV, optical, and near-IR. Even with X-ray surveys where AGNs are most easily detected, obscured AGN can be missed or mistaken for lower-luminosity AGN \citep{Fornasini2018}. There is strong evidence for a large population of obscured AGNs that are masquerading as low-luminosity AGNs \citep{Lambrides2020}.

Some common techniques for identifying AGNs are utilizing IR selection methods. AGNs are typically luminous infrared sources; this IR emission is energy that has been re-emitted by a dusty torus \citep{Treister2012}. A benefit of IR selection is that this re-emission is mainly isotropic, resulting in both unobscured and obscured sources having similar probabilities of detection. Dust obscures a large fraction of star formation and AGN activity in the early Universe, making the infrared spectrum imperative to our studying of these processes. It is possible to identify star formation and AGN signatures in dust emission thanks to the extensive infrared data from space telescopes \citep{Lacy2004, Stern2005, Assef2013, Donley2012, Stern2012, Messias2013, Kirkpatrick2015}. 

Dusty SFGs are also very luminous in the IR and they can outshine the AGN and be contaminants in an IR selection process. High luminosity AGNs are typically easy to detect in the infrared; what is unclear is why some galaxies with IR detections indicative of an AGN lack an corresponding X-ray detection \citep{DelMoro2016}. Are they heavily obscured? Low luminosity? Or is it star formation in exotic environments masquerading as an AGN? A population of objects whose physical nature is ambiguous when looking at their flux and MIR and optical emission was found in the \textit{Chandra} Deep Field \citep{Lambrides2020}. These objects appear as low-luminosity AGNs based on their X-ray luminosities; however if only looking at their MIR and optical line emission, these same objects are classified as moderate to high luminosity AGNs \citep{Lambrides2020}. 

Other common AGN identification techniques are with X-ray selection methods. X-ray emission from AGN is attributed to the up-scattering of UV photons by high energy electrons in the corona just above the accretion disk \citep{Treister2012}. X-ray AGN detection with $\rm log (L_{x})>42$ ${\rm erg}$ ${\rm s}^{-1}$ is unambiguous as they are usually about 1-5 times more luminous than SFGs. However X-ray selection is biased against the most obscured sources due to the strong absorption of their X-ray signal (by factors of 10 - 100) at energies $<$10 keV. Thus detecting heavily obscured and low-luminosity sources requires long integration times.

A significant fraction of AGNs are moderately obscured ($\sim 25\%$ at $N_{H} = 10^{23}\,{\rm cm}^{-2}$) at $z = 2-3$ \citep{Treister2004}. Compton-thick AGN ($N_{H} > 2 \times 10^{24}\,{\rm cm}^{-2}$), can be missed in all but the deepest X-ray surveys \citep{Ballantyne2006, Tozzi2006}. \cite{Gilli2007} predicts with fits to the cosmic X-ray background that at high luminosities ($\rm log(L_{0.5-2keV})>43.5$ ${\rm erg}$ ${\rm s}^{-1}$) both moderately obscured and Compton-thick AGNs are as numerous as unobscured AGNs, and at low luminosities ($\rm log(L_{0.5-2keV})<43.5$ ${\rm erg}$ ${\rm s}^{-1}$) these moderately obscured and Compton-thick AGNs are four times more numerous. The problem with missing these lower-luminosity obscured and Compton thick AGNs when using X-rays alone is that they are not only a significant fraction of the overall AGN population, but also serve as important probes of SMBH/galaxy co-evolution \citep{Donley2012}. 

As X-ray surveys are incomplete, we must use alternate methods that are not sensitive to dust obscuration to identify the presence of an AGN \citep{Kirkpatrick2013}. From the near-IR to the far-IR there are AGN signatures, making infrared color selection techniques a promising method of selecting AGNs that are missed by X-ray surveys \citep{Kirkpatrick2013}. Another method to reveal obscured X-ray sources is to perform X-ray stacking, where images of galaxies with no X-ray signals are stacked to determine if a faint X-ray signal can be found. While X-ray AGN are intrinsically luminous and predominately occur prior to galaxy quenching, the role that X-ray undetected AGN play in galaxy evolution remains uncertain. A census of X-ray detected and X-ray undetected AGN in the distant Universe is needed to develop a framework of how AGN fit into a galaxy’s life cycle.

In this paper, we present a stacking analysis of the X-ray emission from galaxies in the COSMOS field that have an IR detection but are lacking a X-ray detection. This analysis allows us to identify obscured AGN candidates in galaxies currently identified as SFGs. 

We describe our sample in Section \ref{sec:Data}. Our IR selection method is described and verified in section \ref{sec:IR}. We present our X-ray stacking analysis in Section \ref{sec:Stacking}. Obscured AGN candidates and their affect on their host galaxies are presented in Sections \ref{sec:Obscured} and \ref{sec:disc}. Throughout this paper, we assume a standard cosmology with $H_0 = 70$ ${\rm km}$ ${\rm s}^{-1}$ ${\rm Mpc}^{-1}$, $\Omega_{\rm M} = 0.3$, and $\Omega_{\Lambda} = 0.7$

\section{COSMOS Catalog Data Selection\label{sec:Data}}

The sample discussed in this paper is taken from the Cosmic Evolution Survey (COSMOS), which covers a $\rm2\,deg^2$ equatorial field. We use the COSMOS2015 catalog which contains precise photometric redshifts ($z$) and stellar masses ($M_{*}$) for more than half a million objects in the COSMOS field \citep{Laigle_2016}. This near-IR selected catalog is optimized for the study of galaxy evolution and environments in the early Universe as it contains Y-band photometry from Subaru/Hyper-Suprime-Cam and $\rm YJHK_S$ photometry from the UltraVISTA-DR2 survey \citep{Laigle_2016}. The deepest regions reach a 90\% completeness limit of $10^{10}M_\odot$ to $z$ = 4 \citep{Laigle_2016}. 

For this analysis, we select all galaxies in COSMOS2015 with redshifts $0.5 \leq z \leq 3$ (above $z = 3$, the IRAC bands no longer trace dust emission) and $M_{*} > 10^{9.5} M_{\odot}$. This mass and redshift cut resulted in a parent sample of 109,691 galaxies. The following analysis was performed on this parent sample.

\begin{figure}[ht!]
\begin{center}
\includegraphics[width=3.25in,keepaspectratio]{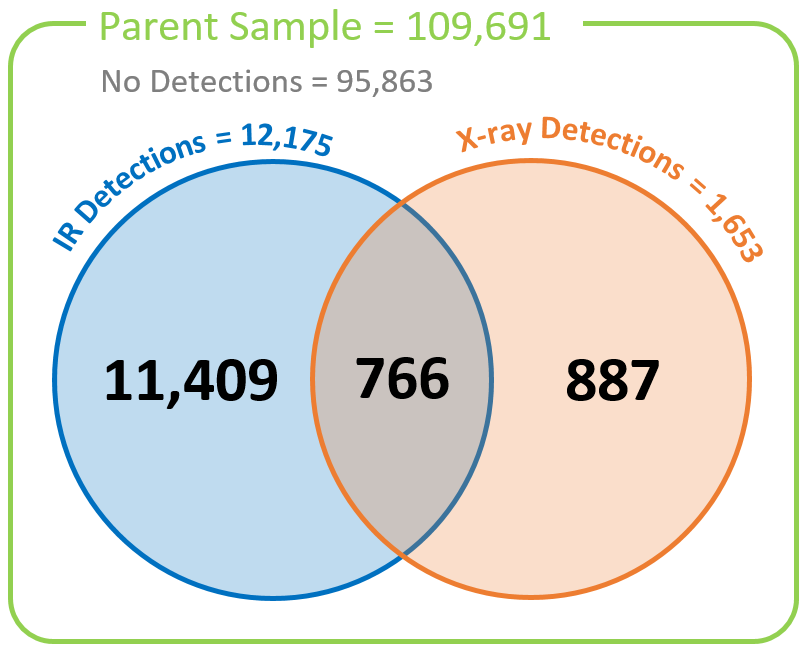}
\caption{COSMOS field detection comparison. \label{fig:Venn}}
\end{center}
\end{figure}

We classify an IR detection as a galaxy with a \textit{Spitzer} $24 \mu m$ detection. \cite{Laigle_2016} obtained their $24 \mu m$ data from the COSMOS MIPS-selected band-merged catalog published by \cite{LeFloch2009}. 12,175 galaxies in our parent sample have an IR detection. We then use these $24 \mu m$ detections to calculate the IR luminosity ($L_{IR}$) for these galaxies using the empirical templates from \cite{Kirkpatrick2015}. These are the IR luminosities referenced throughout this work.

X-ray counterparts for each galaxy are taken from the {\it Chandra}-COSMOS Legacy Survey Multiwavelength Catalog \citep{Marchesi2016}. The X-ray portion of this data was collected during {\it Chandra} Cycle 14, in the {\it Chandra}-COSMOS Legacy Survey \citep{Civano2016}. The flux limits of this survey at 20\% of the area of the whole survey are: Soft Band (0.5-2 keV) = $3.2 \times 10^{-16} {\rm erg\ s}^{-1} {\rm cm}^{-2}$; Hard band (2-10keV) = $2.1 \times 10^{-15} {\rm erg\ s}^{-1} {\rm cm}^{-2}$; Full band (0.5-10 keV) = $1.3 \times 10^{-15} {\rm erg\ s}^{-1} {\rm cm}^{-2} $ \citep{Marchesi2016}. By pairing this X-ray data with our IR data, we created a multiwavelength catalog. We use \textit{Chandra} because it has the deepest, high-resolution data available at these wavelengths. We divide our IR detected galaxies into two categories, IR galaxies that have an X-ray detection (hard X-ray data) and IR galaxies that do not have an X-ray detection. We use hard X-rays an an indicator of detection because soft X-rays (0.5 - 2 keV) are more easily attenuated by dust. Figure \ref{fig:Venn} shows the breakdown of the detection data we have for our sample. As we are only looking at the hard band for this analysis, all X-ray luminosities ($L_{X}$) discussed refer to the hard band luminosities. 

\begin{figure}[ht!]
\begin{center}
\includegraphics[width=3.4in,keepaspectratio]{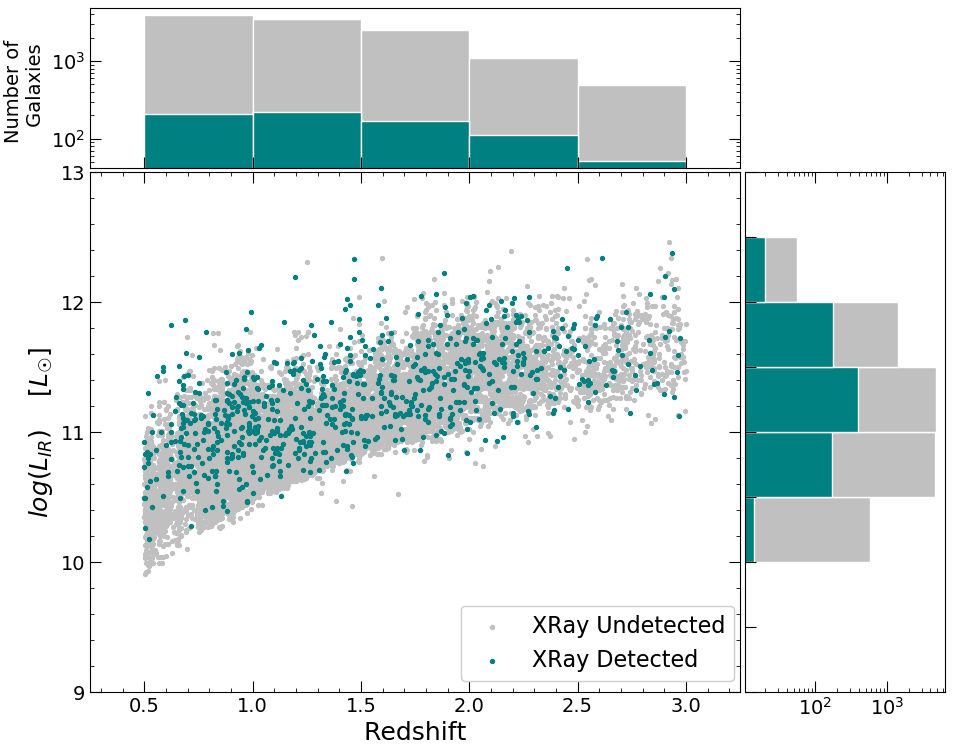}
\caption{Summary of COSMOS field parent sample of infrared detected galaxies. The similar trends between the X-ray detected and undetected sources in  $L_{IR}$ vs $z$ indicate that there is no bias between our X-ray detected (teal points) and non X-ray detected sources (grey points) samples in $L_{IR}$ or $z$. \label{fig:IR_Z}}
\end{center}
\end{figure}

The majority of our IR detected galaxies lack an X-ray detection, and for the galaxies that have X-ray detections over half do not have IR data available. This highlights the need to use innovative X-ray detection techniques such as X-ray stacking to lessen this inequity, resulting in an increased number of multiwavelength detected galaxies. This figure also also suggests the need for IR stacking, with hundreds of galaxies having an X-ray detection but no IR detection. IR stacking techniques are beyond the scope of this work. For the remainder of our analysis we only work with galaxies in the blue circle, those that have an IR detection and may or may not have an X-ray detection. Throughout this work we will refer to these groups as X-ray detected galaxies (galaxies with an IR and a X-ray detection) and X-ray undetected galaxies (galaxies that only have an IR detection). Figure \ref{fig:IR_Z} summarizes the X-ray detection statistics of our IR detected galaxies in COSMOS2015. We find no biases in our X-ray detected and undetected sources with either $L_{IR}$ or $z$.

\section{IR AGN Classification\label{sec:IR}}
\subsection{Mid-IR Color Classification\label{subsec:Color}}
The most prominent IR-AGN selection techniques  \citep[e.g.][]{Donley2012,Lacy2004,Stern2005} are based on \textit{Spitzer}/IRAC colors (typically $S_{8\mu m}/S_{4.5\mu m}$ and $S_{5.8\mu m}/S_{3.6\mu m}$) or WISE colors \citep{Stern2012, Assef2015}, but these are biased toward luminous AGNs that outshine their hosts \citep{Mendez2013}. By including longer wavelength \textit{Herschel} data, this bias can be overcome by comparing the amount of warm dust (heated by the AGN) relative to the amount of cold dust (in the ISM), allowing for the identification of AGNs within star-forming galaxies \citep{Kirkpatrick2013}. The ratio of 250$\mu$m/24$\mu$m flux is lower where AGN heating raises the dust temperature, while 8$\mu$m/3.6$\mu$m separates stellar radiation from power-law AGN radiation. Importantly, this color selection is sensitive to intrinsically luminous but obscured AGNs and to less luminous AGNs in star-forming hosts. 

Utilizing the mid-IR color-color selection technique outlined in \cite{Kirkpatrick2013}, we quantify the fraction of mid-IR (5-15$\mu$m) emission attributable to the dust heating from the AGN, $f({\rm AGN})_{\rm MIR}$. We refer the reader to that paper for details of the color selection method. \cite{Kirkpatrick2017} calculated this $f({\rm AGN})_{\rm MIR}$ for all sources in the COSMOS field, which we use for our IR detected galaxies. This color methodology has an accuracy of $\Delta f({\rm AGN})_{\rm MIR} = 0.3$ \citep{Kirkpatrick2013, Kirkpatrick2017}. For this analysis, we divide our galaxies into four classifications based on their $f({\rm AGN})_{\rm MIR}$ as follows: IR SFGs ($f({\rm AGN})_{\rm MIR} = 0.0$), likely SFGs ($0.0< f({\rm AGN})_{\rm MIR} \leq 0.3$), composites ($0.3 <f({\rm AGN})_{\rm MIR} \leq 0.7$), and IR AGN ($f({\rm AGN})_{\rm MIR} > 0.7$). We classify as obscured AGN candidates a galaxy with $f({\rm AGN})_{\rm MIR} > 0.3$ and no X-ray detection.

\begin{figure}[ht!]
\begin{center}
\includegraphics[width=3.4in,keepaspectratio]{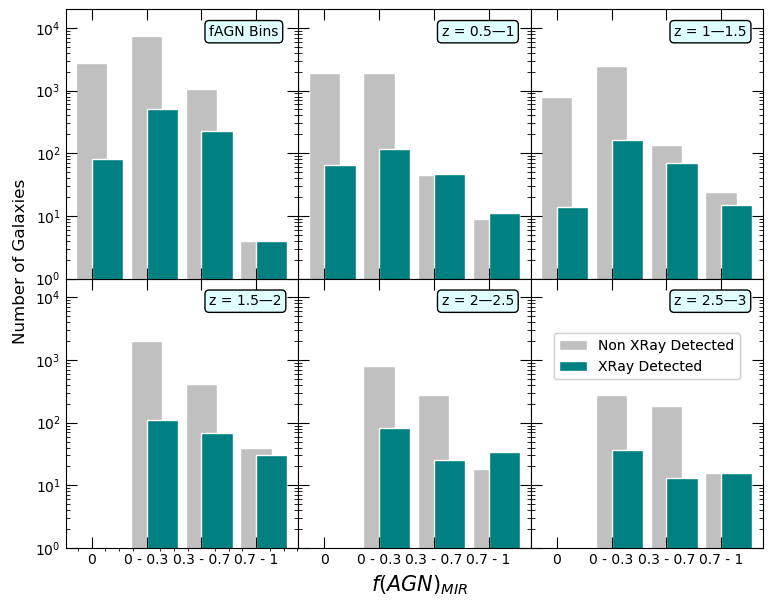}
\caption{We present the X-ray detection statistics of our IR detected galaxies. X-ray detected galaxies (teal) and X-ray undetected galaxies (grey) sorted into bins of $f({\rm AGN})_{\rm MIR}$. The upper left panel is the full sample only binned by $f({\rm AGN})_{\rm MIR}$ at all redshifts. Subsequent panels show the sample broken down into increasing redshift bins. \label{fig:XNoX}}
\end{center}
\end{figure}

The distribution of $f({\rm AGN})_{\rm MIR}$ for our X-ray detected and undetected galaxies are shown in Figure \ref{fig:XNoX}. We see a similar trend in quantities of X-ray undetected galaxies as we do in the quantities of X-ray detected galaxies, with respect to $f({\rm AGN})_{\rm MIR}$, both peaking at $f({\rm AGN})_{\rm MIR} = 0 - 0.3$ and then decreasing as $f({\rm AGN})_{\rm MIR}$ increases. This peak at $f({\rm AGN})_{\rm MIR} = 0 - 0.3$ is expected as our X-ray data from \textit{Chandra} is deep, allowing for the detection of many faint galaxies. Of these X-ray undetected galaxies, a large portion have a $f({\rm AGN})_{\rm MIR}$ indicative of the presence of an AGN ($f(AGN)_{MIR} > 0.3$). The X-ray detected sample is biased towards galaxies with larger $f({\rm AGN})_{\rm MIR}$ fractions when compared with the X-ray undetected sample. This indicates the necessity of pairing IR and X-ray data to gain a full census of the AGN population.

\subsection{Mid-IR vs Alternate Detection Methods\label{subsec:MAPHYS}}

\begin{figure}[ht!]
\begin{center}
\includegraphics[width= 3.4in,keepaspectratio]{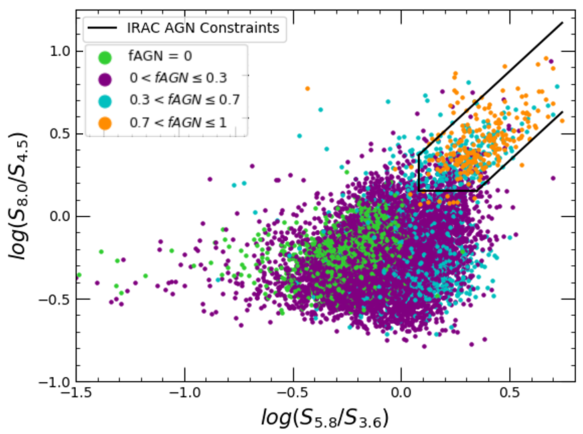}
\caption{We plot IR detected galaxies from our sample in IRAC color space with the criteria of \cite{Donley2012} (black lines). Each galaxy is color-coded with the mid-IR color selection technique outlined in \cite{Kirkpatrick2017}. The mid-IR color selection technique selects more AGN candidates than the restrictive IRAC color selection.\label{fig:IRAC}}
\end{center}
\end{figure}

Our color selection technique selects more AGN candidates than the more restrictive IRAC-only criteria of \cite{Donley2012}. Figure \ref{fig:IRAC} shows our IR detected galaxies from COSMOS2015, color-coded by $f({\rm AGN})_{\rm MIR}$ from \cite{Kirkpatrick2017}, plotted with \cite{Donley2012} selection criteria. \cite{Donley2012} identifies AGN as galaxies that fall within these black lines. It can be seen that the color selection technique we use identifies AGNs ($f({\rm AGN})_{\rm MIR} > 0.7$, orange) outside of these IRAC criteria. \cite{Donley2012} took the best fits of the median SEDs of {\it XMM-Newton}-selected Quasi Stellar Object (QSOs) that fell both inside and outside of their selection criteria, and they found that both Type 1 and Type 2 QSOs that are missed by their IRAC criteria have slightly redder UV-optical continuua and prominent 1.6 $\mu$m stellar bumps. This indicates that the IRAC selection method is most likely to miss lower-luminosity AGN with luminous hosts, and this omission is elevated if the AGN emission is obscured \citep{Donley2012}.

We also compared this mid-IR color selection $f({\rm AGN})_{\rm MIR}$ with the $f({\rm AGN})$ calculated with MAGPHYS spectral energy decomposition by \cite{Chang2017}. MAGPHYS (Multi-wavelength Analysis of Galaxy Physical Properties) is a self-contained model package used to interpret observed spectral energy distributions of galaxies in terms of galaxy-wide physical parameters \citep{Cunha2008}. \cite{Chang2017} uses a custom version of the MAGPHYS code, with an AGN component introduced to the fitting, to provide a public catalog that could include AGN information. See \cite{Chang2017} for a description of this modification.

\begin{figure}[ht]
\begin{center}
    \includegraphics[width=3.4in,keepaspectratio]{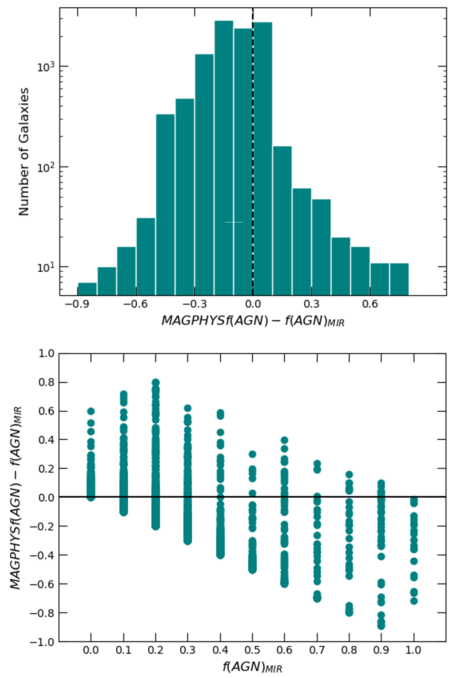}
    \caption{Galaxies were selected that had both a $f({\rm AGN})_{\rm MIR}$ value from \cite{Kirkpatrick2017} and a MAGPHYS $f({\rm AGN})$ value from \cite{Chang2017}. These values were then subtracted from each other (MAGPHYS $f({\rm AGN})$ - $f({\rm AGN})_{\rm MIR}$) in order to determine how well these methods agree for each galaxy. \textit{\textbf{Top:}} Distribution of differences between MAGPHYS $f({\rm AGN})$ and $f({\rm AGN})_{\rm MIR}$. \textit{\textbf{Bottom:}} MAGPHYS $f({\rm AGN})$ - $f({\rm AGN})_{\rm MIR}$ as a function of $f({\rm AGN})_{\rm MIR}$.  MAGPHYS tends to have larger $f({\rm AGN})_{\rm MIR}$ values at low $f({\rm AGN})_{\rm MIR}$, whereas color selected $f({\rm AGN})_{\rm MIR}$ values are larger at higher $f({\rm AGN})_{\rm MIR}$ \label{fig:MAPHYS}}
\end{center}
\end{figure}

We compare our color selection with MAGPHYS as spectral energy distribution (SED) decomposition is becoming an increasingly popular tool for estimating $f({\rm AGN})_{\rm MIR}$ \citep[e.g.][]{yang2021}. Particularly, this tool has already been applied to all of the COSMOS galaxies in \citet{Chang2017}. Color selection is an important technique in its own right, since it can save valuable time over SED decomposition when applied to large data sets. We compare our results with MAGPHYS to demonstrate how well color selection alone can quantify $f({\rm AGN})_{\rm MIR}$. Figure \ref{fig:MAPHYS} showcases this comparison.

The distribution of $f({\rm AGN})_{\rm MAGPHYS}-f({\rm AGN})_{\rm MIR}$, shown in the top panel, has a mean of $-0.12$ and a standard deviation of $-0.16$. The top panel highlights that this mid-IR color selection is in strong agreement with the MAGPHYS analysis, with only a handful of galaxies that have a difference in $f({\rm AGN})$ values greater than $0.5$. This distribution is skewed due to many of our $f({\rm AGN})_{\rm MIR}$ values being greater than their MAGPHYS counterparts. The bottom panel displays how the $f({\rm AGN})_{\rm MAGPHYS}-f({\rm AGN})_{\rm MIR}$ changes with $f({\rm AGN})_{\rm MIR}$. MAGPHYS tends to have larger $f({\rm AGN})_{\rm MIR}$ values at low $f({\rm AGN})_{\rm MIR}$, whereas color selection $f({\rm AGN})_{\rm MIR}$ values are larger at higher $f({\rm AGN})_{\rm MIR}$. These are important results as color selection is faster than SED decomposition, and therefore may be the preferred method for large surveys.

\section{X-ray Stacking and Analysis \label{sec:Stacking}}
\subsection{X-ray Stacking Details}
With the aim of probing obscured/low-luminosity AGNs below the sensitivity threshold of the \textit{Chandra} COSMOS survey, we perform X-ray stacking analysis using the \textit{Chandra} stacking tool CSTACK \footnote{CSTACK(http://lambic.astrosen.unam.mx/cstack/) was developed by Takamitsu Miyaji.} v4.32 \citep{Miyaji2008}. CSTACK provides the net (background $-$ subtracted) count rate and count rate errors in the soft (0.5–2keV) and hard (2–8keV) bands for each target, using all 117 observations from the \textit{Chandra} COSMOS-Legacy survey and associated exposure maps \citep{Civano2016}. The hard band used in CSTACK is not as wide as the hard band of our Chandra data (2-10 keV); all stacked galaxies use hard band data from 2-8 keV, while data for detected sources is 2-10 keV. This discrepancy between energy ranges of stacked and detected galaxies will not hinder our results as the Chandra effective area falls rapidly at high energies, resulting in little difference between these two energy ranges.

This survey is a mosaic with a high level of overlap. This allows for an object to be observed multiple times at different off-axis angles and have a very uniform PSF of about 2" \citep{Civano2016}. The \textit{Chandra} PSF has variation with off-axis angles, therefore for each observation of an object, CSTACK defines a circular source extraction region with its size determined by the 90\% encircled counts fraction (ECF) radius ($r_{90}$) (with a minimum of 1"). This allows for the signal-to-noise ratios of the stacked signals to be optimized. Each object is in the center of a $30\times30$ arcsec$^2$ area background. A 7"-radius circle around the object is excluded from the background as well as circles around any other detected X-ray sources in this area. By default, CSTACK only includes observations in which the source is located withing 8' of the aim point where $r_{90}<7"$. 

Due to the computation time needed for large stacks with CSTACK, it was not efficient to submit every combination of bins we analyzed to CSTACK. Instead, we ran all X-ray undetected sources through CSTACK as one large stack. This resulted in a master file of the source data, (including source counts, source exposure time, and background counts) for each of our X-ray undetected galaxies that is used to calculate a net count-rate (weighted average) for a stack. We then pulled the source data for individual galaxies for each stack as needed. This allowed us to calculate a net count-rate, weighted by exposure times, for each bin combination in a fraction of the time CSTACK uses. 

\begin{figure*}[ht!]
\begin{center}
\includegraphics[width=6in,keepaspectratio]{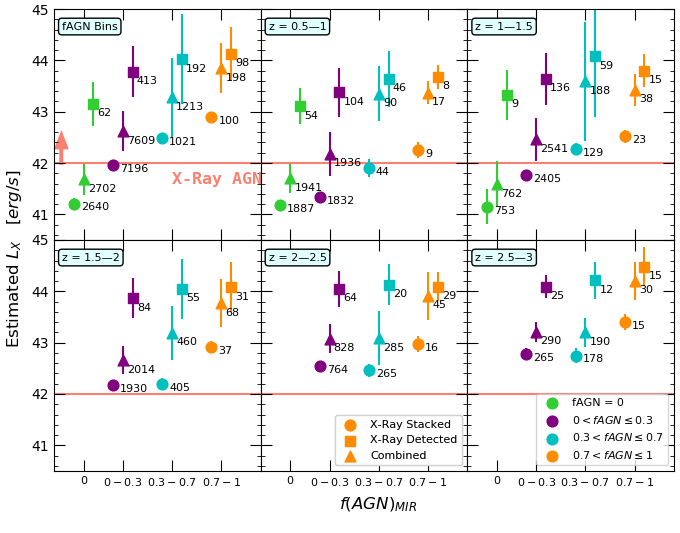}
\caption{$L_X$ as a function of $f({\rm AGN})_{\rm MIR}$ of stacked galaxies (circles), X-ray detected galaxies (squares), and the combined weighted average of the two (triangles). All classifications are binned in terms of $f({\rm AGN})_{\rm MIR}$ (colors) an $z$. The upper left panel is galaxies only binned in terms of $f({\rm AGN})_{\rm MIR}$ at all redshifts, with subsequent panels are increasing with $\Delta z = 0.5$. Numbers indicate the number of galaxies in each bin. IR SFGs have a $f(AGN)_{MIR}=0$ (green). Most stacked bins all show elevated $L_X$ values above IR SFGs. As redshift increases we see an increase in $L_X$ of stacked bins. Points that are absent are $f(AGN)_{MIR}=0$ at $z > 1.5$, this is due to the lack of galaxies in those bins at those redshifts. Colors definitions remain consistent throughout this paper for these bins.\label{fig:LXs}}
\end{center}
\end{figure*}

Using this net count-rate we calculated a flux for each bin using the PIMMS function in the Chandra Proposal Planning Toolkit. PIMMS is a tool used for estimating the source count rate or flux for a specific mission. It makes this estimate from the flux in a specified energy bound, or a count rate estimated from a previous mission. We input to PIMMS the net count-rate calculated for each bin, using a power law model, a galactic $N_H$ value of $26 \times 10^{20} \rm cm^{-2}$, and a photon index of 1.4, to replicate the model used in \cite{Fornasini2018}. We refer the reader to that work for a full description of model choice. To verify that a photon index of 1.4 was valid for this research we calculated fluxes for all stacks at different photon index values from $1.4-2.2$ and found there was minimal change in flux with change in photon index. $\Gamma= 1.4$ because is commonly used in AGN studies as it is the photon index of the X-ray background \citep{DeLuca2004}, and it is the photon index used for the spectral model in \cite{Fornasini2018}, which we aim to replicate. With these parameters in PIMMS we received an absorbed flux as an output that was used to calculate the observed $L_{X}$.

We compare our stacked bins to X-ray detected galaxies divided into the same bins. There is a possibility that the X-ray detected galaxies themselves are highly obscured. To investigate this, we calculated the obscuration corrected luminosities for the X-ray detected sample. The {\it Chandra}-COSMOS Legacy Survey provides observed $L_{X}$ values. This survey also included the luminosity absorption correction for both the soft and hard band. The absorption correction values were pulled from the Chandra-COSMOS Legacy Survey Multiwavelength Catalog. We applied the absorption correction terms to the luminosities of individual X-ray detected galaxies, then we took the mean of the absorption corrected $L_{X}$ values ($L_{X}^{Corr}$), with standard deviation errors. The $L_{X}^{Corr}$ values show only a slight increase over the observed $L_{X}$ values. This indicated that there are little to no obscured AGNs in our X-ray detected sample. Moving forward, we use only the observed, uncorrected $L_{X}$ values for the X-ray detected sample 
in order to accurately compare to the observed $L_{X}$ values we get from stacking the X-ray undetected sources. These observed and corrected $L_{X}$ values for the X-ray detected sample can be seen in Table \ref{tab:Table1} in the Appendix.

\subsection{X-ray Stacking Results \label{sec:Stacking_Results}}

Figure \ref{fig:LXs} shows the $L_{X}$ values resulting from our X-ray stacking. Our galaxies are binned in terms of $f({\rm AGN})_{\rm MIR}$, with an $f({\rm AGN})_{\rm MIR} < 0.3$ representing SFGs (green = IR SFGs and purple = likely SFGs), as well as separated into redshift bins of size $\Delta z = 0.5$. Our X-ray detected galaxies are divided into the same bins and their mean $L_{X}$ and standard deviations are plotted. In order to analyze how $L_{X}$ changes with $f({\rm AGN})_{\rm MIR}$ without the biases introduced by X-ray sensitivity limits, we combine the X-ray stacked and X-ray detected galaxies in each bin and take a weighted average of their luminosities to get a combined $L_{X}$. In general, $L_{X}$ increases with $f({\rm AGN})_{\rm MIR}$. All stacked bins have lower luminosities than the corresponding X-ray detected bins. Interestingly, most stacked bins have elevated $L_{X}$ values above the IR SFGs. As redshift increases we see an increase in $L_{X}$ in the stacked bins. This is natural given the IR detection limit of COSMOS. At higher redshift, we are only capable of detecting high luminosity sources (Figure \ref{fig:IR_Z}). There are no $f(AGN)_{MIR}=0$ points at $z > 1.5$ which is due to the lack of galaxies in those bins at those redshifts. The uncertainty of the photometric redshifts is not expected to be a dominant source of uncertainty in our results when we divide sources into redshifts bins of $\Delta z = 0.5$. Luminosities and redshifts for all bins are listed in Table \ref{tab:Table2.1} \& \ref{tab:Table2.2} in the Appendix \ref{sec:Append}.

The high SFRs of dusty galaxies in this epoch could contaminate X-ray detections due to other X-ray emitting mechanisms that occur in highly star forming systems such as X-ray binaries (XRBs). \cite{Basu-Zych2013} found that the 2-10 keV X-ray luminosity of their X-ray stacked galaxies evolves weakly with redshift and SFR, and that this redshift evolution is driven by metallicity evolution in high mass XRBs. Many subsequent works have found evidence to support this relationship between the X-ray luminosity of XRBs, redshift, and SFR \citep{Lehmer2016, Aird2018, Fornasini2019,Fornasini2020}. It is possible that the elevated $L_{X}$ signals in our IR detected galaxies are due to XRBs. We calculated the expected $L_{X}$ due to these binaries using the scaling relation found by \cite{Lehmer2010}, as follows:

\begin{equation}
    \rm L^{gal}_{HX} = \alpha M_{*} + \beta SFR
\end{equation}

\begin{figure}[ht!]
    \begin{center}
    \includegraphics[width=3.4in,keepaspectratio]{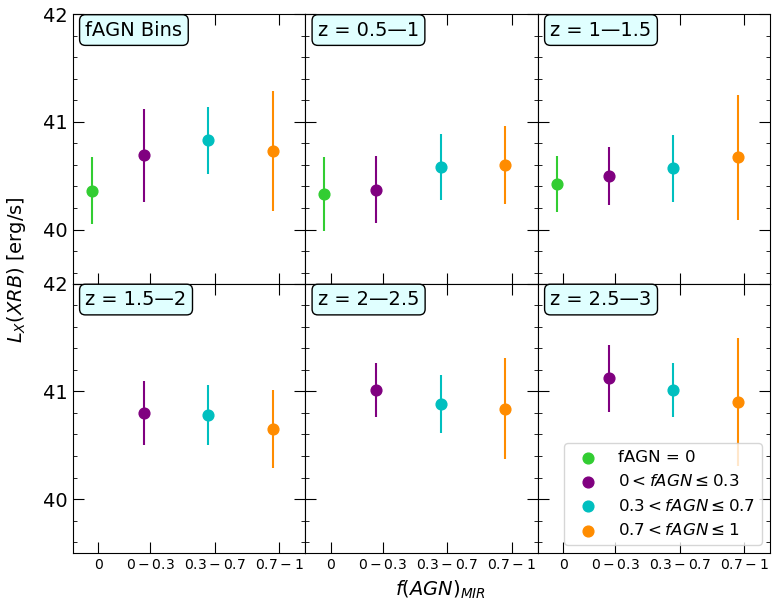}
    \caption{Calculated X-ray luminosity from XRBs in each stacked bin. These $f({\rm AGN})_{\rm MIR}$ bins (colors) were additionally divided into redshift bins of size $\Delta z = 0.5$. There is no discernible difference in $L_{X}$ from XRBs, not only in comparison to the IR-SFGs but also between each of the bins, at all redshifts. This indicates that the elevated $L_{X}$ signals over the IR SFGs (\ref{fig:LXs}) is not caused by XRBs. \label{fig:XRB}}
    \end{center}
\end{figure}

with $\alpha=(9.05\pm0.37)\times10^{28}$ erg s$^{-1}$ M$^{-1}_{\odot}$ and $\beta=(1.62\pm0.22)\times10^{39}$ erg s$^{-1}$ $(\rm M_{\odot}\,\rm yr^{-1})^{-1}$. Our results (Figure \ref{fig:XRB}) show no discernible difference in $L_{X}$ from XRBs, not only in comparison to the IR SFGs but also between each of the bins, at all redshifts. We conclude XRBs are not the major cause of these elevated $L_{X}$ signals (\ref{fig:LXs}) over the IR SFGs.

An increase in $L_{X}$ with $f({\rm AGN})_{\rm MIR}$ could be a side-effect of mass bias if the higher $f({\rm AGN})_{\rm MIR}$ bins are skewed towards more massive galaxies. More massive galaxies would presumably have more fuel available for feeding the SFR and the SMBH, producing larger luminosities. We analyze how $L_{X}$ changes with mass only in the low $f({\rm AGN})_{\rm MIR}$ bins, as these are the only bins with enough sources to split into multiple mass bins and still produce statistically relevant stacks with CSTACK. Working with $f(AGN)_{MIR} = 0 - 0.3$, we bin our stacked galaxies by stellar mass with $\Delta M_{\ast} = 0.5$.

\begin{figure}[ht!]
    \begin{center}
    \includegraphics[width= 3.35in,keepaspectratio]{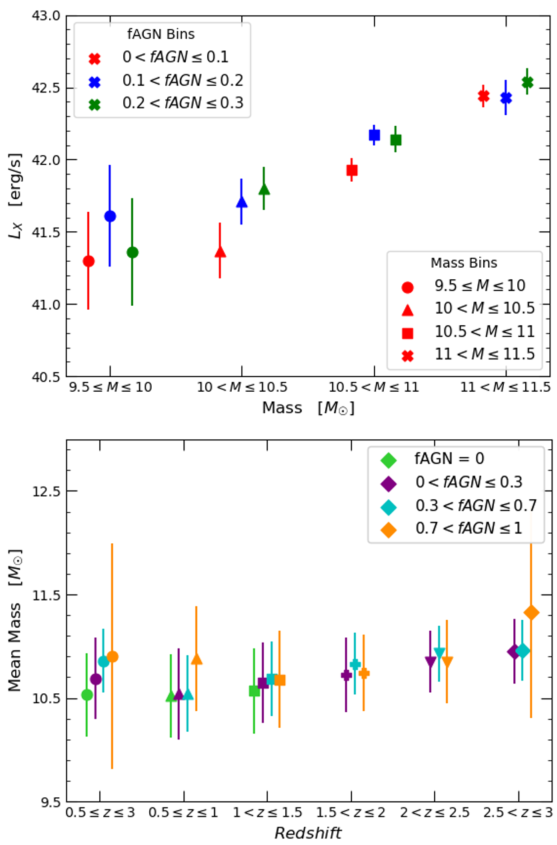}
    \caption{Investigation into the effect of mass and $f({\rm AGN})_{\rm MIR}$ on $L_{X}$. \textit{\textbf{Top:}} $f({\rm AGN})_{\rm MIR}$ bins (colors) are additionally divided into mass bins with $\Delta M_{\ast} = 0.5$ plotted at their mean $M_{\ast}$. $L_{X}$ is more strongly correlated with $M_{\ast}$. \textit{\textbf{Bottom:}} Further investigating the effect of mass, continued at all redshifts. All $f({\rm AGN})_{\rm MIR}$ bins additionally divided by redshift ( $\Delta z = 0.5$) (shapes) and plotted at their mean $M_{\ast}$. In each redshift bin, all $f({\rm AGN})_{\rm MIR}$ bins have similar mean masses, indicating that the observed trend between $L_{X}$ and $f({\rm AGN})_{\rm MIR}$ is not driven by an underlying relationship with $M_{\ast}$. Points that are absent are $f(AGN)_{MIR}=0$ at $z > 1.5$ due to the lack of galaxies in those bins at those redshifts. \label{fig:Mass}}
    \end{center}
\end{figure}

\begin{figure*}[ht!]
\begin{center}
\includegraphics[width=6in,keepaspectratio]{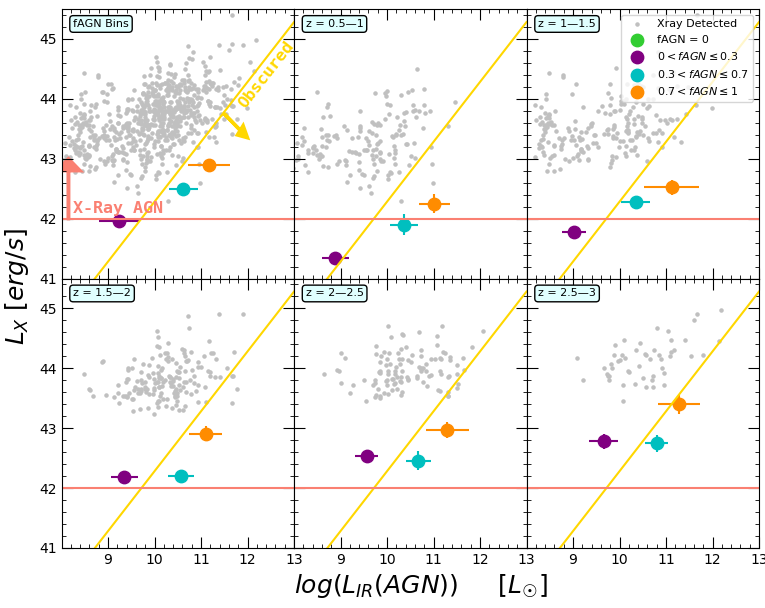}
\caption{Hard X-ray luminosities vs total IR AGN contribution of X-ray detected (grey) and X-ray stacked galaxies (colors). The yellow line indicates the cut off for a galaxy to be considered obscured as defined in \cite{Chang2017}. All stacks with $f({\rm AGN})_{\rm MIR} > 0.3$ qualify as obscured and have $L_{X}$ signals that indicate a luminous AGN ($\rm log L_{X} > 42$ ergs/s) at all redshifts, classifying the galaxies in these stacks as obscured AGN candidates. This plot is comparable to \cite{Chang2017}. Based on the location of these stacked bins, it is clear that with X-ray stacking we probing a regime that is currently lacking in data. Points that are absent are all $f(AGN)_{MIR}=0$ bins, this is due to this $f(AGN)_{MIR}$ bin having an IR AGN contribution of zero. \label{fig:OBS}}
\end{center}
\end{figure*}

The top panel of Figure \ref{fig:Mass} depicts our investigation into the effect of mass and $f({\rm AGN})_{\rm MIR}$ on $L_{X}$ of our stacked galaxies. It is clear that changes in $L_{X}$ are correlated more strongly with mass than with $f({\rm AGN})_{\rm MIR}$, allowing for larger $f({\rm AGN})_{\rm MIR}$ bins to be used throughout this analysis. This is emphasized by the bottom panel where we analyze the mean mass, the driver of $L_{X}$, of each of our stacked $f({\rm AGN})_{\rm MIR}$ bins. Each $f({\rm AGN})_{\rm MIR}$ bin has the same mean mass and similar standard deviations, in each redshift bin. Consistency in mass distributions between $f({\rm AGN})_{\rm MIR}$ bins means that mass is not responsible for driving the observed trend between $L_{X}$ and $f({\rm AGN})_{\rm MIR}$.

We also compared the mean specific SFR ($sSFR=SFR/M_\ast$) of all the bins at each redshift and found no discernible difference between the bins nor at different redshifts, ruling out that high $f({\rm AGN})_{\rm MIR}$ bins have higher sSFRs. This is further evidence that there is a physical mechanism other than star formation that is causing these $L_{X}$ signals.

\begin{figure*}[ht!]
    \begin{center}
    \includegraphics[width= 7.20in,keepaspectratio]{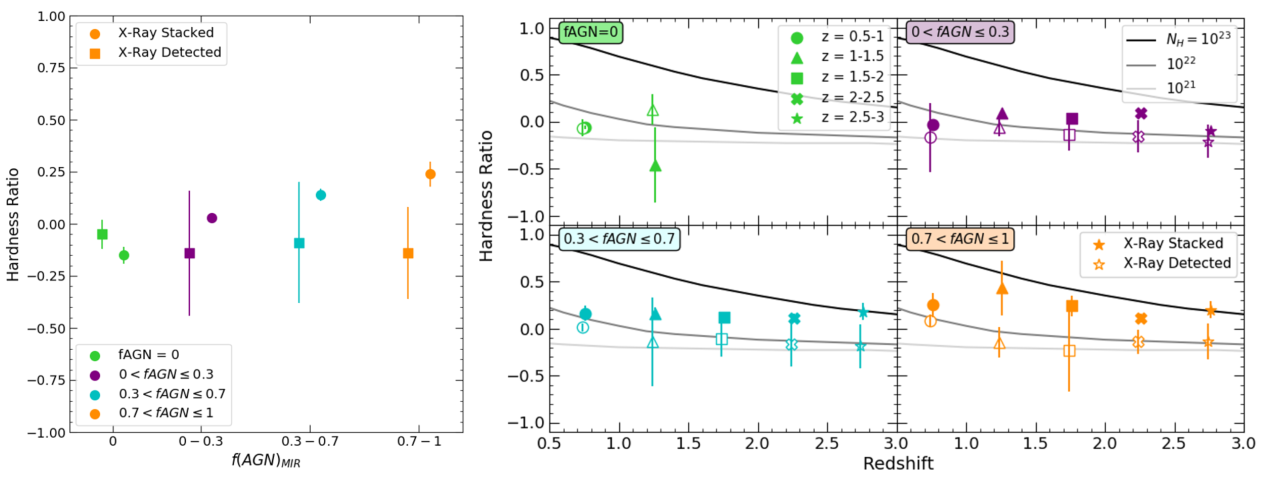}
    \caption{\textit{\textbf{Left:}} Hardness ratios of stacked bins (circles) and X-ray detected bins (squares) broken into bins of $f({\rm AGN})_{\rm MIR}$ (colors). For stacked galaxies, bins with a higher $f({\rm AGN})_{\rm MIR}$ have a higher hardness ratio indicating that there is more gas present at higher $f({\rm AGN})_{\rm MIR}$ than with the IR SFGs ($f({\rm AGN})_{\rm MIR}$ = 0). \textit{\textbf{Right:}} The bins from the left plot are further broken into bins of redshift with $\Delta z = 0.5$ (shapes). Each panel is a different $f({\rm AGN})_{\rm MIR}$, also indicated by color. The lines indicate the hardness ratios expected for different absorbed power-law model with a photon index of 1.4 as done in \cite{Fornasini2020}; different color-density of the lines indicate different column densities. All AGNs here are Compton thin ($N_{H}=10^{22} cm^{-2}$). We see a consistent separation between the stacked (solid) and detected (open) galaxies with $f({\rm AGN})_{\rm MIR} > 0.7$ (orange) in both panels. For $0.3 < f({\rm AGN})_{\rm MIR} < 0.7$ there is a slight increase in obscuration as $z$ increases. X-ray detected bins have no difference in HR with changes in redshift. Points that are absent are $f(AGN)_{MIR}=0$ at $z > 1.5$, this is due to the lack of galaxies in those bins at those redshifts. \label{fig:HARD}}
    \end{center}
\end{figure*}

\section{Obscured AGN Candidates \label{sec:Obscured}}

For this study we define obscuration as the ratio between IR AGN and X-ray luminosities, ${L_{IR}(AGN)}/{L_{X}}$. This relation was initially computed for $L_{6\mu m}/L_{X}$ in \cite{Perna2015}, and \cite{Chang2017} demonstrated that $L_{6}$ is a large enough fraction of ${L_{IR}(AGN)}$ for ${L_{IR}(AGN)}/{L_{X}}$ to also work as a means to identify potentially heavily obscured AGN. $L_{IR}(AGN)$ was calculated from $L_{IR}$ by multiplying the $L_{IR}$ of a galaxy or stacked bin by the corresponding $f({\rm AGN})_{\rm Total}$ value. We use the relation $f({\rm AGN})_{\rm Total}$ = $0.66 \times f({\rm AGN})_{\rm MIR}^2-0.35 \times f({\rm AGN})_{\rm MIR}$ from \citet{Kirkpatrick2015}. $L_{IR}$ is the IR luminosity, calculated by taking the template from \cite{Kirkpatrick2015} that corresponds to the $f({\rm AGN})_{\rm MIR}$ of each source, scaled to 24$\mu$m and integrated from 8-1000 $\mu m$. $L_{X}$ is the observed hard X-ray luminosity. For stacked bins, $L_{IR}$ is the mean IR luminosity of the galaxies in the bin. We classify a galaxy or a stacked bin as obscured when $L_{IR}/L_{X} > 20 $ as defined in \cite{Chang2017}. This definition was empirically chosen in \cite{Chang2017} according to their Compton-thick AGNs. An $L_{IR}/L_{X} > 20 $ corresponds to a $N_{H}=10^{24} {\rm cm^{-2}}$ \citep{Perna2015}.

Figure \ref{fig:OBS} summarizes our obscuration analysis. Not all of our stacked bins lie below the obscured line, nor at all redshifts. Bins with the $f({\rm AGN})_{\rm MIR} > 0.3$ have $L_{X}$ signals that indicate luminous AGN ($L_{X} > 10^{42}$ ${\rm erg\, s}^{-1}$) and are below the obscured line, classifying the galaxies in these stacks as obscured AGN candidates. Based on the locations of our stacked bins in comparison to the X-ray detected galaxies, they highlight that with X-ray stacking we are able to probe a regime that is currently unexplored. Figure \ref{fig:OBS} accentuates a key difference in our stacked galaxies. Galaxies with an $f({\rm AGN})_{\rm MIR} > 0.3$ are predominantly obscured AGN according to ${L_{IR}}/{L_{X}}$, while galaxies with $0 < f({\rm AGN})_{\rm MIR} < 0.3$ are not. At $z < 1.5$, these galaxies may be star forming ($L_{X} < 10^{42}$ ${\rm erg s}^{-1}$), but at $z > 1.5$, they fall securely in the $L_{X}$ range of an AGN. These galaxies seem then to be lower luminosity AGN in strongly star forming hosts-- the exact type of AGN not targeted with classic IR selection techniques \citep{Donley2012, Assef2013}. A potential follow up study could directly test this using a field with deeper \textit{Chandra} data such as GOODS-S or CDF-S. 

Hardness ratios (HRs) are a key tool in analysing the possible presence of gas and dust in a system. For each bin of both stacked and X-ray detected galaxies we calculate the hardness ratio using the count rate in the soft (S, 0.5-2 keV) and the hard (H, 2-8 keV for stacked, 2-10 keV for detected) bands \citep{Fornasini2018}. The HR for each stack is defined as $({H - S})/({H + S})$. We ensured that we only uses galaxies in each bin that have both soft and hard data for accurate calculation of HRs. Soft X-rays get more easily attenuated by gas and dust; therefore stacks with higher HRs indicate that there is more gas and dust in the system. 

Figure \ref{fig:HARD} displays the hardness ratios for all bins. The left plot shows the effect of $f({\rm AGN})_{\rm MIR}$ on HRs for both stacked and X-ray detected sources. We see that for the stacked bins, as $f({\rm AGN})_{\rm MIR}$ increases so does HR, echoing the conclusions drawn from Figure \ref{fig:OBS}. There is a significant increase in HRs over the stacked IR-SFGs. Typically, stacked bins have the same HR as corresponding detected bins, except at $f({\rm AGN})_{\rm MIR} > 0.7$. The right plot takes the points from the left plot and further divides them into redshift bins of $\Delta z = 0.5$. 

We compare our hardness ratios with the expected hardness ratio of a galaxy, assuming a simple absorbed power-law spectrum with $\Gamma=1.4$. We caution the reader, however, that the absorbed power-law model might be too simplistic for these sources, and that X-ray reflection can contribute significantly in obscured AGN. Additionally, a steeper photon index would result in a different correlation between $N_H$ and hardness ratio. All AGNs here are Compton thin ($N_{H}=10^{22} {\rm cm^{-2}}$), and therefore have some obscuring gas around them. We do not see a significant different in HRs in stacks with $0 < f({\rm AGN})_{\rm MIR} < 0.7$ (purple and blue), not only between redshift bins but also between stacked and X-ray detected galaxies, although there is a slight increase in obscuration in the stacked $0.3 < f({\rm AGN})_{\rm MIR} < 0.7$ points (blue). The separation between the HRs of stacked and X-ray detected galaxies with $f({\rm AGN})_{\rm MIR} > 0.7$ (orange) is consistent with the left plot. X-ray detected bins have no difference in HR with changes in redshift in all $f({\rm AGN})_{\rm MIR}$ bins. 

We posit that the sources with $0 < f({\rm AGN})_{\rm MIR} < 0.3$ are a mix of SFGs, low luminosity AGN, and obscured AGN. Deeper X-ray observations or the increased IR wavelength coverage of JWST will enable a more robust measurement of the AGN bolometric luminosity, so that stacking is not necessary. Similarly, the difference between the $f({\rm AGN})_{\rm MIR} > 0.3$ and the $f({\rm AGN})_{\rm MIR} > 0.7$ sources seems to be driven by X-ray luminosity rather than obscuration as the ${L_{IR}}/{L_{X}}$ ratio is the same. This indicates that the IR can be an effective tools for isolating less luminous and obscured AGN in the distant Universe.

\section{Discussion}
\label{sec:disc}
\subsection{Possible Interpretation of Timescale}

If we make the assumption that our sources are drawn from the same population, but are in different evolutionary stages, we can use information from Figure \ref{fig:LXs} to estimate the percentage of their life AGN spend in each bin/phase. These assumptions are supported by the fact that all of our stacked and detected galaxies have similar stellar masses and star formation rates as seen in Figure \ref{fig:MS}. On the basis of these two properties, we can assume we are looking at galaxies drawn from the same parent population. However in order to form stronger conclusions, we would need to also be able to accurately measure black hole masses and gas reservoirs, allowing us to dive deep into the evolution and growth timescales. For now, we present a speculative scenario that we can test with future work.

Our toy evolutionary model is that galaxies start as an SFG, and then some process triggers the fueling of the central SMBH. The SMBH grows in luminosity, which correlates with the $f({\rm AGN})_{\rm MIR}$ bin. If we assume AGN move through each phase from $f({\rm AGN})_{\rm MIR} < 0.3$ to $0.3 <f({\rm AGN})_{\rm MIR} < 0.7$, and finally $f({\rm AGN})_{\rm MIR} > 0.7$, we can use the number of galaxies in each bin to find the percentage of time spent in each bin/phase. Although fueling of AGN is a stochastic process, with AGN flickering on timescales of 10,000 years \citep{Hickox2014}, our sample is large enough that we can still examine timescales in a statistical sense. 

\begin{figure}[ht!]
    \begin{center}
    \includegraphics[width=3.3in,keepaspectratio]{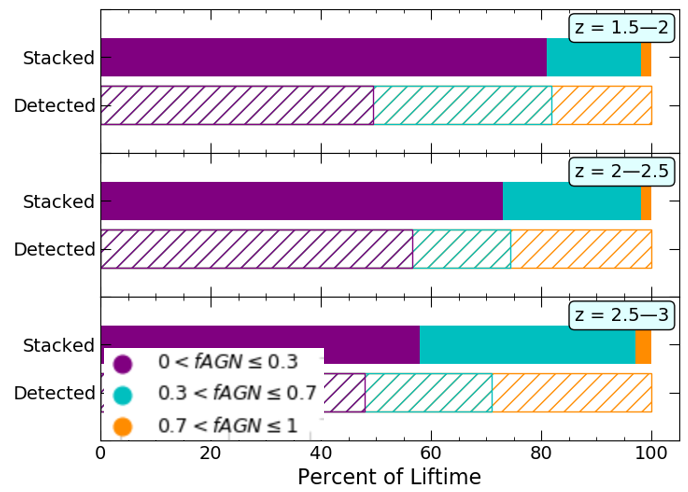}
    \caption{We present the percentage of an AGN's life time spent in each bin/phase, $f({\rm AGN})_{\rm MIR} < 0.3$ (purple), $0.3 <f({\rm AGN})_{\rm MIR} < 0.7$ (teal), and  $f({\rm AGN})_{\rm MIR} > 0.7$ (orange), for X-ray stacked (solid) and X-ray detected (hashed). Each panel increases with a $\Delta z = 0.5$ from top to bottom. Percentages were calculated based on number of galaxies in each bin. Both X-ray stacked bins and X-ray detected bins spend the largest amount of time in the $f({\rm AGN})_{\rm MIR} < 0.3$ phase. \label{fig:Timescale}}
    \end{center}
\end{figure}

These timescale estimates are summarized in Figure \ref{fig:Timescale}. The percentage of time spent in each phase was calculated for X-ray stacked and X-ray detected galaxies in bins of $f({\rm AGN})_{\rm MIR}$ and $\Delta z = 0.5$. We only look at $z \geq 1.5$ because at these redshifts all bins have $L_{X} > 10^{42}\, {\rm erg\,s}^{-1}$ ensuring the X-ray luminosity is mainly attributable to AGN, with fewer SFG interlopers (see Figure \ref{fig:LXs}). Our stacked bins consistently have the longest time spent in the $f({\rm AGN})_{\rm MIR} < 0.3$ phase and the least amount of time spent in the $f({\rm AGN})_{\rm MIR} > 0.7$. Although as $z$ increases, the time spent in the $f({\rm AGN})_{\rm MIR} < 0.3$ phase decreases and the time spent in the $0.3 < f({\rm AGN})_{\rm MIR} < 0.7$ and $f({\rm AGN})_{\rm MIR} > 0.7$ phases increase. X-ray detected galaxies also have the most time spent in the $f({\rm AGN})_{\rm MIR} < 0.3$ phase at all redshifts. As $z$ increases the time spent in the $f({\rm AGN})_{\rm MIR} < 0.3$ decreases, and the time spent in the $0.3 <f({\rm AGN})_{\rm MIR} < 0.7$ and $f({\rm AGN})_{\rm MIR} > 0.7$ phases increase, with switch between $0.3 <f({\rm AGN})_{\rm MIR} < 0.7$ and $f({\rm AGN})_{\rm MIR} > 0.7$ phases as the higher redshifts. 
This indicates that the majority of an AGN's lifetime is spent in a lower luminosity phase which may be missed by X-ray telescopes. Furthermore, $f({\rm AGN})_{\rm MIR} < 0.7$ are missed by IRAC color selection and WISE color selection, which are tuned for the brightest AGN \citep{Donley2012,Assef2013}. For a complete picture of AGN growth, sensitive IR detection is required. The upcoming JWST will greatly enable the detection of less luminous AGN hidden within star forming galaxies \citep{Kirkpatrick2017}.

\subsection{AGN on the Main Sequence \label{sec:MS}}
We now turn our attention to the evolution of the host galaxy itself. We look at the evolution of the host galaxy with regards to the main sequence -- the tight correlation between SFR and $M_\ast$ that holds for the majority of galaxies. The main sequence for each redshift bin was calculated 
with the relationship between $M_\ast$ and SFR parameterized in \cite{Lee2015} at $z = 1.2$:

\begin{equation}
    \rm log(SFR_{MS}) = 1.72 - log\left[ 1+ \left( \frac{M_\ast}{2 \times 10^{10} M_\odot}\right)^{-1.07} \right]
\end{equation}

\begin{flushleft}
We then normalized depending on redshift \citep{Speagle2014}:
\end{flushleft}

\vspace{-0.25 in}

\begin{equation}
\label{eqn:MS}
    \rm SFR_{MS}(\textit{z}) = \left( \frac{1+\textit{z}}{1+1.2}\right)^{2.9} \times \rm SFR_{MS}(\textit{z} = 1.2)
\end{equation}

\begin{figure*}[ht!]
    \begin{center}
    \includegraphics[width=6in,keepaspectratio]{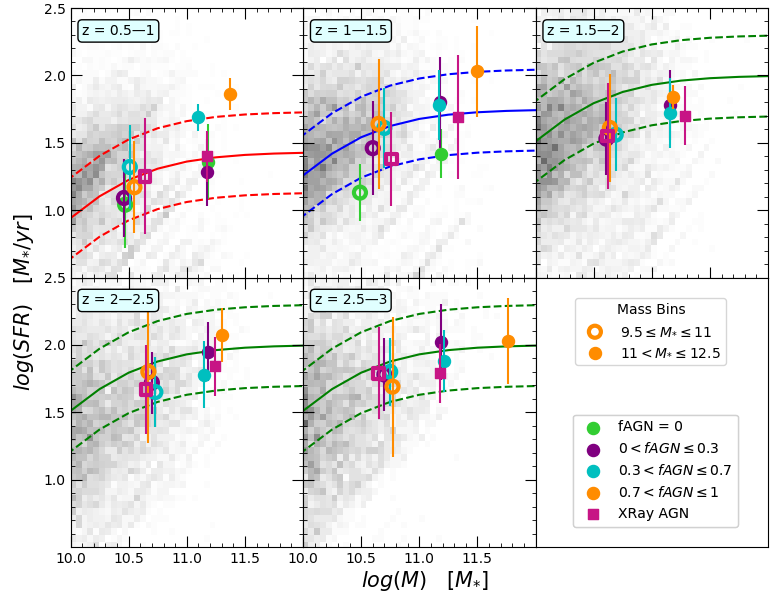}
    \caption{Main Sequence Analysis: SFR mass sequence for COSMOS galaxies. Panels indicate binning of $\Delta z = 0.5$. Solid colored lines are main sequence lines with dashed lines indicating the $0.3$ $dex$ error. Colors of solid lines change with redshift, except after $z > 1.5$. The grey contours are the full COSMOS2015 catalog. Stacked (circles) and X-ray detected (squares) bins are additionally broken into into low-mass (open points) and high-mass (solid points), and then plotted at the mean $M_{*}$ of the bin. The points that are not on the plot due to lack of data are: all $f({\rm AGN})_{\rm MIR} = 0$ at $z > 1.5$. All stacked bins lie within $0.3$ $dex$ of the main sequence line, except for $f({\rm AGN})_{\rm MIR} > 0.7$ at $z = 0.5-1$. Our stacked bins are indistinguishable from X-ray AGN at all redshifts. \label{fig:MS}}
    \end{center}
\end{figure*}

The main sequence was calculated for five different redshift bins from $0.5 \leq z \leq 3$ with $\Delta z = 0.5$. However, we found that at $z>2$, Equation \ref{eqn:MS} was no longer consistent with the full COSMOS15 sample. In Figure \ref{fig:MS}, we use the $z=1.5$ main sequence in the higher redshift bins, as that is consistent with the full COSMOS population.

In order to determine if our obscured AGN candidates correlate with their host's SFR, we calculate each bin's SFR using their $L_{IR}$ and $L_{IR}(AGN)$ values as follows: 

\begin{equation}
    \rm SFR\, [M_\odot/yr]= ( L_{IR} - L_{IR}(AGN) ) \times (1.59\times 10^{-10})
\end{equation}

\begin{figure}[ht!]
    \begin{center}
    \includegraphics[width=3.35in,keepaspectratio]{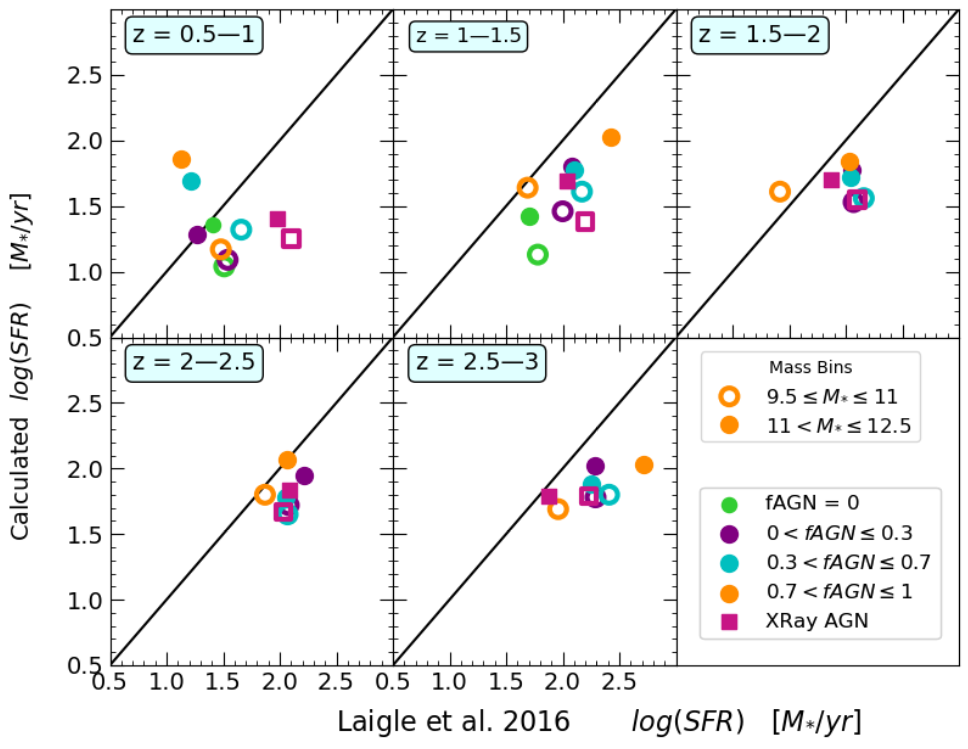}
    \caption{Comparison between our calculated SFRs and the SFRs provided in COSMOS2015. COSMOS2015 SFRs are consistently higher at all redshifts, except for the points in bins $f({\rm AGN})_{\rm MIR} < 0.3$ at $z = 0.5-1$. Points that are absent are $f(AGN)_{MIR}=0$ at $z > 1.5$, this is due to the lack of galaxies in those bins at those redshifts. \label{fig:SFR_Comp}}
    \end{center}
\end{figure}

This relationship was also used to calculate the SFR of the X-ray detected bins as well. We calculate our SFRs as oppose to using the SFRs provided in COSMOS2015 because the SFRs in COSMOS2015 are not corrected for AGN contribution. We compare our calculated SFRs with those in the COSMOS2015 catalog binned in the same way and find that the COSMOS2015 SFRs are consistently larger than our calculated SFRs, as would be expected for they include the AGN component. This comparison can be seen in Figure \ref{fig:SFR_Comp}.

Figure \ref{fig:MS} displays our main sequence analysis. The mean SFR was used for both stacked bins and X-ray AGN bins. All bins were additionally divided into a low mass bin and a high mass bin. The main sequence for each redshift bin was calculated with their $0.3$ $dex$ spread. Colors of solid lines change with redshift, except after $z > 1.5$. The galaxies in COSMOS do not evolve much, therefore higher redshifts were fit with the main sequence from $z = 1.5 - 2$. We show the full COSMOS2015 catalog in grey. For the grey points, SFR was calculated from UV/optical SED fitting.

The IR-detected sources lie within $0.3$ $dex$ of the main sequence line at each redshift, except for $f({\rm AGN})_{\rm MIR} > 0.7$ at $z = 0.5-1$. Furthermore, at $z>1$, the SFRs of all $f({\rm AGN})_{\rm MIR}$ bins are indistinguishable from each other, and evolve very little with mass. Most importantly, our stacked bins are indistinguishable from X-ray AGN at all redshifts. This argues against X-ray detected AGN being found only in a special time in a galaxy's life, right before quenching. Our results rather support the picture that SFR and AGN luminosity are not tied together in any particular way, similar to results from other stacking analyses \citep{Stanley2015}.

\section{Conclusions}
We uncover obscured and lower luminosity AGN in the COSMOS field using an X-ray stacking technique paired with IR color selection methods. The combination of these techniques allowed us to measure the energetics of these obscured AGN, and compare their host galaxies with X-ray bright AGN. Our main results are as follows:

1. This work illustrates that stacking can reveal low luminosity AGN as well as obscured AGN that are currently undetected. X-ray binaries and mass bias have been ruled out as a major source of X-ray luminosity in our stacked bins. 

2. Stacked bins with $f({\rm AGN})_{\rm MIR} < 0.3 $ appear to host low luminosity AGN as indicated by their hardness ratio and by the ratio of ${L_{IR}(AGN)}/{L_{X}}$. These plots also indicate bins with $f({\rm AGN})_{\rm MIR} > 0.3 $ have obscured AGN.

3. Obscured AGN are found in galaxies where star formation is still ongoing. X-ray stacked AGNs are indistinguishable from X-ray detected AGNs on the main sequence. 

4. We can use these results to predict how long AGN last in each phase. Assuming AGN go through $f({\rm AGN})_{\rm MIR} < 0.3$, to $0.3 <f({\rm AGN})_{\rm MIR} < 0.7$, and end in $f({\rm AGN})_{\rm MIR} > 0.7$, we can use their relative quantity in each bin to estimate the timescale of each phase. At all relevant redshifts, AGN spend the largest portion of their life in the $f({\rm AGN})_{\rm MIR} < 0.3$ phase.

5. The analysis of the combination of obscuration and our toy evolution model suggest that the fraction of time an AGN spends in an obscured phase may increase with redshift.

\section{Acknowledgements}
We would like to thank the referee for their thoughtful comments which have helped to improve the clarity and impact of this work. This is work is based in part on observations made with \textit{Spitzer Space Telescope}, operated by the Jet Propulsion Laboratory, California Institute of Technology under a contract with NASA, \textit{Herschel Space Observatory}, a European Space Agency Cornerstone Mission with significant participation by NASA, and archival data from \textit{Chandra X-ray Observatory} operated by the Smithsonian Astrophysical Observatory for and on behalf of NASA. C.H. acknowledges support from NASA through Chandra Archival Grant Award Number 21700372 issued by the Chandra X-ray Center. and through the NASA FINESST Award Number 19-ASTRO20-0078.
\textit{Software} CSTACK \citep{Miyaji2008}

\section{Appendix \label{sec:Append}}

\begin{deluxetable*}{cccccc}
\tabletypesize{\footnotesize}
\tablecolumns{6}
\tablewidth{1.0\columnwidth}
\tablenum{1}
\tablecaption{Summary of Absorption Corrected X-Ray Luminosities \label{tab:Table1}}

\tablehead{
\colhead{$f({\rm AGN})_{\rm MIR}$} \vspace{-0.2cm} & 
\colhead{Num}&
\colhead{$z$}&
\colhead{$log(L_{X})$} & 
\colhead{$log(L_{X}^{Corr}$} &
\colhead{$NH$ $(10^{22})$}\\ 
\vspace{-0.2cm}&&&
\colhead{${\rm erg}$ ${\rm s}^{-1}$}&
\colhead{${\rm erg}$ ${\rm s}^{-1}$}&
\colhead{${\rm cm}^{-2}$}
}
\colnumbers
\decimalcolnumbers
\startdata
0 & 39 & all & 43.19$\pm$0.45 & 43.22$\pm$0.45 & 3.69$\pm$4.59\\
0 & 34 & 0.5-1 & 43.14$\pm$0.36 & 43.17$\pm$0.36 & 3.78$\pm$4.79\\
0 & 5 & 1-1.5 & 43.43$\pm$0.49 & 43.46$\pm$0.50 & 3.05$\pm$2.76\\
0-0.3 & 327 & all & 43.83$\pm$0.47 & 43.85$\pm$0.47 & 7.43$\pm$18.32\\
0-0.3 & 84 & 0.5-1 & 43.44$\pm$0.45 & 43.46$\pm$0.45 & 2.16$\pm$12.06\\
0-0.3 & 107 & 1-1.5 & 43.69$\pm$0.48 & 43.73$\pm$0.47 & 6.98$\pm$8.42\\
0-0.3 & 60 & 1.5-2 & 43.92$\pm$0.39 & 43.92$\pm$0.40 & 6.70$\pm$24.13\\
0-0.3 & 55 & 2-2.5 & 44.09$\pm$0.33 & 44.11$\pm$0.34 & 12.23$\pm$25.19\\
0-0.3 & 21 & 2.5-3 & 44.13$\pm$0.20 & 44.16$\pm$0.20 & 20.32$\pm$24.85\\
0.3-0.7 & 170 & all & 44.05$\pm$0.86 & 44.07$\pm$0.86 & 6.61$\pm$15.91\\
0.3-0.7 & 40 & 0.5-1 & 43.68$\pm$0.51 & 43.72$\pm$0.50 & 4.61$\pm$4.84\\
0.3-0.7 & 54 & 1-1.5 & 44.11$\pm$1.16 & 44.13$\pm$1.15 & 4.74$\pm$16.66\\
0.3-0.7 & 48 & 1.5-2 & 44.08$\pm$0.56 & 44.08$\pm$0.61 & 6.70$\pm$19.65\\
0.3-0.7 & 17 & 2-2.5 & 44.18$\pm$0.37 & 44.20$\pm$0.37 & 10.65$\pm$13.35\\
0.3-0.7 & 11 & 2.5-3 & 44.25$\pm$0.34 & 44.27$\pm$0.33 & 16.31$\pm$18.76\\
0.7-1 & 89 & all & 44.15$\pm$0.50 & 44.18$\pm$0.48 & 13.18$\pm$19.80\\
0.7-1 & 8 & 0.5-1 & 43.68$\pm$0.23 & 43.73$\pm$0.24 & 5.02$\pm$5.40\\
0.7-1 & 13 & 1-1.5 & 43.85$\pm$0.29 & 43.88$\pm$0.26 & 8.66$\pm$8.90\\
0.7-1 & 27 & 1.5-2 & 44.11$\pm$0.48 & 44.14$\pm$0.46 & 11.58$\pm$19.04\\
0.7-1 & 28 & 2-2.5 & 44.10$\pm$0.27 & 44.13$\pm$0.30 & 14.85$\pm$20.32\\
0.7-1 & 13 & 2.5-3 & 44.53$\pm$0.35 & 44.56$\pm$0.33 & 22.44$\pm$28.04\\
\enddata

\tablecomments{X-ray luminosities and corrected X-ray luminosities for X-ray detected galaxies. We list the average $L_X$, $L_X^{Corr}$ (absorption corrected luminosity), and $N_H$. These values for individual galaxies are from \citet{Lambrides2020}.}\end{deluxetable*}

\begin{deluxetable*}{lcccccccc}
\tabletypesize{\footnotesize}
\tablecolumns{9}
\tablewidth{0pt} 
\tablenum{2.1}
\tablecaption{Summary of Stacking Data \label{tab:Table2.1}}

\tablehead{
\colhead{Model} \vspace{-0.2cm} &
\colhead{$f({\rm AGN})_{\rm MIR}$} & 
\colhead{Num}&
\colhead{$z$}&
\colhead{$log(L_{X})$} & 
\colhead{$log(L_{IR})$} &
\colhead{$log(L_{IR}(AGN))$} &
\colhead{SFR} &
\colhead{$log(Mass)$}\\ 
\vspace{-0.2cm}&&&&
\colhead{${\rm erg}$ ${\rm s}^{-1}$}&
\colhead{$L_{\odot}$}&
\colhead{$L_{\odot}$}&
\colhead{$M_{*}/yr$}&
\colhead{$M_{\odot}$}}
 
\colnumbers
\decimalcolnumbers
\startdata
{Stacked}& 0 & 2640 & all & 41.21$\pm$0.10 & 10.88$\pm$0.30 & 0$\pm$0 & 12.18$\pm$8.30 & 10.53$\pm$0.40\\
{Stacked}& 0 & 1887 & 0.5 - 1 & 41.18$\pm$0.10 & 10.86$\pm$0.33 & 0$\pm$0 & 11.40$\pm$7.31 & 10.57$\pm$0.40\\
{Stacked}& 0 & 753 & 1 - 1.5 & 41.15$\pm$0.34 & 10.95$\pm$0.22 & 0$\pm$0 & 14.13$\pm$0.40 & 10.57$\pm$0.41\\
{Stacked}& 0-0.3 & 7196 & all & 41.96$\pm$0.03 & 11.25$\pm$0.43 & 9.23$\pm$0.43 & 27.82$\pm$27.46 & 10.69$\pm$0.39\\
{Stacked}& 0-0.3 & 1832 & 0.5-1 & 41.34$\pm$0.08 & 10.90$\pm$0.29 & 8.89$\pm$0.29 & 12.62$\pm$8.54 & 10.54$\pm$0.44\\
{Stacked}& 0-0.3 & 2405 & 1-1.5 & 41.77$\pm$0.07 & 11.04$\pm$0.25 & 9.02$\pm$0.25 & 17.09$\pm$10.03 & 10.65$\pm$0.39\\
{Stacked}& 0-0.3 & 1930 & 1.5-2 & 42.18$\pm$0.07 & 11.37$\pm$0.29 & 9.35$\pm$0.29 & 36.55$\pm$24.34 & 10.72$\pm$0.36\\
{Stacked}& 0-0.3 & 764 & 2-2.5 & 42.54$\pm$0.08 & 11.58$\pm$0.25 & 9.56$\pm$0.25 & 59.36$\pm$34.23 & 10.85$\pm$0.30\\
{Stacked}& 0-0.3 & 265 & 2.5-3 & 42.78$\pm$0.12 & 11.68$\pm$0.31 & 9.66$\pm$0.31 & 54.87$\pm$54.87 & 10.95$\pm$0.31\\
{Stacked}& 0.3-07 & 1021 & all & 42.49$\pm$0.06 & 11.45$\pm$0.31 & 10.62$\pm$0.31 & 38.00$\pm$26.78 & 10.86$\pm$0.31\\
{Stacked}& 0.3-07 & 44 & 0.5-1 & 41.90$\pm$0.17 & 11.20$\pm$0.30 & 10.37$\pm$0.30 & 21.66$\pm$15.08 & 10.54$\pm$0.37\\
{Stacked}& 0.3-07 & 129 & 1-1.5 & 42.27$\pm$0.11 & 11.18$\pm$0.31 & 10.35$\pm$0.31 & 20.41$\pm$14.37 & 10.69$\pm$0.36\\
{Stacked}& 0.3-07 & 405 & 1.5-2 & 42.40$\pm$0.11 & 11.40$\pm$0.28 & 10.57$\pm$0.28 & 33.80$\pm$21.65 & 10.83$\pm$0.30\\
{Stacked}& 0.3-07 & 265 & 2-2.5 & 42.46$\pm$0.16 & 11.50$\pm$0.27 & 10.67$\pm$0.27 & 42.61$\pm$26.42 & 10.93$\pm$0.27\\
{Stacked}& 0.3-07 & 178 & 2.5-3 & 42.75$\pm$0.14 & 11.63$\pm$0.25 & 10.80$\pm$0.25 & 57.46$\pm$32.58 & 10.96$\pm$0.29\\
{Stacked}& 0.7-1 & 100 & all & 42.90$\pm$0.07 & 11.52$\pm$0.48 & 11.17$\pm$0.48 & 28.82$\pm$32.02 & 10.90$\pm$1.09\\
{Stacked}& 0.7-1 & 9 & 0.5-1 & 42.25$\pm$0.16 & 11.37$\pm$0.33 & 11,02$\pm$0.33 & 20.42$\pm$15.52 & 10.88$\pm$0.51\\
{Stacked}& 0.7-1 & 23 & 1-1.5 & 42.52$\pm$0.13 & 11.47$\pm$0.59 & 11.12$\pm$0.59 & 26.01$\pm$35.35 & 10.68$\pm$0.47\\
{Stacked}& 0.7-1 & 37 & 1.5-2 & 42.91$\pm$0.12 & 11.45$\pm$0.36 & 11.10$\pm$0.36 & 24.58$\pm$20.12 & 10.74$\pm$0.37\\
{Stacked}& 0.7-1 & 16 & 2-2.5 & 42.97$\pm$0.14 & 11.65$\pm$0.47 & 11.30$\pm$0.47 & 39.05$\pm$42.61 & 10.85$\pm$0.40\\
{Stacked}& 0.7-1 & 15 & 2.5-3 & 43.40$\pm$0.16 & 11.63$\pm$0.45 & 11.28$\pm$0.45 & 37.68$\pm$39.37 & 11.33$\pm$1.02\\
\hline
{Detected}& 0 & 62 & all & 43.15$\pm$0.42 & 11.01$\pm$0.23 & 0$\pm$0 & 16.35$\pm$8,61 & 10.87$\pm$0.29\\
{Detected}& 0 & 54 & 0.5-1 & 43.11$\pm$0.35 & 11.00$\pm$0.24 & 0$\pm$0 & 16.07$\pm$8.97 & 10.87$\pm$0.29\\
{Detected}& 0 & 9 & 1-1.5 & 43.33$\pm$0.49 & 11.05$\pm$0.14 & 0$\pm$0 & 18.00$\pm$5.79 & 10.84$\pm$0.29\\
{Detected}& 0-0.3 & 413 & all & 43.78$\pm$0.49 & 11.37$\pm$0.40 & 9.36$\pm$0.40 & 37.31$\pm$34.49 & 10.87$\pm$0.37\\
{Detected}& 0-0.3 & 104 & 0.5-1 & 43.38$\pm$0.48 & 11.12$\pm$0.40 & 9.11$\pm$0.40 & 20.91$\pm$19.03 & 10.82$\pm$0.37\\
{Detected}& 0-0.3 & 136 & 1-1.5 & 43.64$\pm$0.51 & 11.20$\pm$0.33 & 9.18$\pm$0.33 & 25.09$\pm$19.32 & 10.81$\pm$0.34\\
{Detected}& 0-0.3 & 84 & 1.5-2 & 43.87$\pm$0.39 & 11.45$\pm$0.36 & 9.43$\pm$0.36 & 43.90$\pm$36.65 & 10.89$\pm$0.36\\
{Detected}& 0-0.3 & 64 & 2-2.5 & 44.05$\pm$0.35 & 11.63$\pm$0.26 & 9.61$\pm$0.26 & 67.10$\pm$40.80 & 10.98$\pm$0.39\\
{Detected}& 0-0.3 & 25 & 2.5-3 & 44.09$\pm$0.22 & 11.67$\pm$0.24 & 9.65$\pm$0.24 & 73.55$\pm$40.93 & 10.89$\pm$0.35\\
{Detected}& 0.3-07 & 192 & all & 44.02$\pm$0.88 & 11.46$\pm$0.42 & 10.62$\pm$0.42 & 38.69$\pm$37.16 & 10.90$\pm$0.75\\
{Detected}& 0.3-07 & 46 & 0.5-1 & 43.64$\pm$0.54 & 11.20$\pm$0.37 & 10.37$\pm$0.37 & 21.71$\pm$18.38 & 10.78$\pm$0.32\\
{Detected}& 0.3-07 & 59 & 1-1.5 & 44.08$\pm$1.19 & 11.48$\pm$0.51 & 10.65$\pm$0.51 & 40.99$\pm$48.37 & 11.00$\pm$1.00\\
{Detected}& 0.3-07 & 55 & 1.5-2 & 44.05$\pm$0.58 & 11.44$\pm$0.29 & 10.60$\pm$0.29 & 36.96$\pm$25.00 & 10.80$\pm$0.44\\
{Detected}& 0.3-07 & 20 & 2-2.5 & 44.13$\pm$0.40 & 11.65$\pm$0.21 & 10.82$\pm$0.21 & 60.57$\pm$29.67 & 11.00$\pm$0.33\\
{Detected}& 0.3-07 & 12 & 2.5-3 & 44.22$\pm$0.36 & 11.67$\pm$0.33 & 10.84$\pm$0.33 & 63.94$\pm$48.76 & 11.01$\pm$0.28\\
{Detected}& 0.7-1 & 98 & all & 44.13$\pm$0.51 & 11.62$\pm$0.44 & 11.27$\pm$0.44 & 36.96$\pm$37.70 & 10.96$\pm$0.71\\
{Detected}& 0.7-1 & 8 & 0.5-1 & 43.68$\pm$0.23 & 11.29$\pm$0.41 & 10.94$\pm$0.41 & 17.29$\pm$16.15 & 10.71$\pm$0.35\\
{Detected}& 0.7-1 & 15 & 1-1.5 & 43.80$\pm$0.32 & 11.54$\pm$0.45 & 11.20$\pm$0.45 & 30.81$\pm$32.18 & 11.08$\pm$0.87\\
{Detected}& 0.7-1 & 31 & 1.5-2 & 44.08$\pm$0.49 & 11.59$\pm$0.36 & 11.24$\pm$0.36 & 33.82$\pm$27.85 & 10.97$\pm$0.83\\
{Detected}& 0.7-1 & 29 & 2-2.5 & 44.09$\pm$0.28 & 11.64$\pm$0.36 & 11.29$\pm$0.36 & 38.21$\pm$31.57 & 10.98$\pm$0.38\\
{Detected}& 0.7-1 & 15 & 2.5-3 & 44.48$\pm$0.38 & 11.82$\pm$0.47 & 11.47$\pm$0.47 & 57.66$\pm$62.34 & 10.84$\pm$0.34\\
\enddata
\tablecomments{Basic properties of our stacked, X-ray detected, and combined bins.}
\end{deluxetable*}

\begin{deluxetable*}{lccccccccB}
\tabletypesize{\footnotesize}
\tablecolumns{9}
\tablewidth{0pt} 
\tablenum{2.2}
\tablecaption{Summary of Stacking Data cont. \label{tab:Table2.2}}

\tablehead{
\colhead{Model} \vspace{-0.2cm} &
\colhead{$f({\rm AGN})_{\rm MIR}$} & 
\colhead{Num}&
\colhead{$z$}&
\colhead{$log(L_{X})$} & 
\colhead{$log(L_{IR})$} &
\colhead{$log(L_{IR}(AGN))$} &
\colhead{SFR} &
\colhead{$log(Mass)$}\\ 
\vspace{-0.2cm}&&&&
\colhead{${\rm erg}$ ${\rm s}^{-1}$}&
\colhead{$L_{\odot}$}&
\colhead{$L_{\odot}$}&
\colhead{$M_{*}/yr$}&
\colhead{$M_{\odot}$}}
 
\colnumbers
\decimalcolnumbers
\startdata
{Combined}& 0 & 2702 & all & 41.68$\pm$0.31 & 10.89$\pm$0.29 & 0$\pm$0 & 12.27$\pm$8.31	& 10.55$\pm$0.40\\
{Combined}& 0 & 1941 & 0.5-1 & 41.70$\pm$0.28 & 10.86$\pm$0.32 &0$\pm$0 & 11.53$\pm$8.55 & 10.53$\pm$0.39\\
{Combined}& 0 & 762 & 1-1.5 & 41.59$\pm$0.44 & 10.95$\pm$0.22 & 0$\pm$0 & 14.18$\pm$7.30 & 10.57$\pm$0.41\\
{Combined}& 0-0.3 & 7609 & all & 42.62$\pm$0.39 & 11.26$\pm$0.43 & 9.24$\pm$0.43 & 28.33$\pm$27.84 & 10.70$\pm$0.39\\
{Combined}& 0-0.3 & 1936 & 0.5-1 & 42.18$\pm$0.43 & 10.92$\pm$0.30 & 8.90$\pm$0.30 & 13.06$\pm$9.10 & 10.56$\pm$0.43\\
{Combined}& 0-0.3 & 2541 & 1-1.5 & 42.46$\pm$0.42 & 11.05$\pm$0.26 & 9.03$\pm$0.26 & 17.52$\pm$10.52 & 10.66$\pm$0.39\\
{Combined}& 0-0.3 & 2014 & 1.5-2 & 42.66$\pm$0.28 & 11.37$\pm$0.29 & 9.35$\pm$0.29 & 36.86$\pm$24.86 & 10.73$\pm$0.36\\
{Combined}& 0-0.3 & 828 & 2-2.5	 & 43.08$\pm$0.28 & 11.58$\pm$0.25 & 9.56$\pm$0.25 & 59.96$\pm$34.74 & 10.86$\pm$0.31\\
{Combined}& 0-0.3 & 290 & 2.5-3 & 43.21$\pm$0.19 & 11.68$\pm$0.31 & 9.66$\pm$0.31 & 75.52$\pm$53.6 & 10.94$\pm$0.31\\
{Combined}& 0.3-0.7 & 1213 & all & 43.28$\pm$0.77 & 11.45$\pm$0.32 & 10.62$\pm$0.32 & 38.11$\pm$28.42 & 10.87$\pm$0.39\\
{Combined}& 0.3-0.7 & 90 & 0.5-1 & 43.35$\pm$0.53 & 11.20$\pm$0.34 & 10.37$\pm$0.34 & 21.69$\pm$16.77 & 10.68$\pm$0.34\\
{Combined}& 0.3-0.7 & 188 &	1-1.5 & 43.59$\pm$1.16 & 11.30$\pm$0.40 & 10.47$\pm$0.40 & 26.87$\pm$25.04 & 10.81$\pm$0.67\\
{Combined}& 0.3-0.7 & 460 & 1.5-2 & 43.19$\pm$0.52 & 11.40$\pm$0.28 & 10.57$\pm$0.28 & 34.18$\pm$22.05 & 10.83$\pm$0.32\\
{Combined}& 0.3-0.7 & 285 & 2-2.5 & 43.09$\pm$0.35 & 11.51$\pm$0.26 & 10.68$\pm$0.26 & 43.87$\pm$26.65 & 10.93$\pm$0.27\\
{Combined}& 0.3-0.7 & 190 & 2.5-3 & 43.20$\pm$0.28 & 11.63$\pm$0.25 & 10.80$\pm$0.25 & 57.87$\pm$33.60 & 10.96$\pm$0.29\\
{Combined}& 0.7-1 & 198 & all & 43.85$\pm$0.49 & 11.57$\pm$0.46 & 11.22$\pm$0.46 & 32.85$\pm$34.83 & 10.93$\pm$0.89\\
{Combined}& 0.7-1 & 17 & 0.5-1 & 43.37$\pm$0.22 & 11.33$\pm$0.36 & 10.98$\pm$0.36 & 18.95$\pm$15.82 & 10.81$\pm$0.45\\
{Combined}& 0.7-1 & 38 & 1-1.5 & 43.43$\pm$0.31 & 11.50$\pm$0.53 & 11.15$\pm$0.53 & 27.90$\pm$34.10 & 10.88$\pm$0.72\\
{Combined}& 0.7-1 & 68 & 1.5-2 & 43.77$\pm$0.47 & 11.52$\pm$0.36 & 11.17$\pm$0.36 & 28.80$\pm$23.64 & 10.86$\pm$0.64\\
{Combined}& 0.7-1 & 45 & 2-2.5 & 43.91$\pm$0.27 & 11.64$\pm$0.40 & 11.29$\pm$0.40 & 38.51$\pm$35.50 & 10.94$\pm$0.39\\
{Combined}& 0.7-1 & 30 & 2.5-3 & 44.21$\pm$0.37 & 11.73$\pm$0.46 & 11.38$\pm$0.46 & 47.67$\pm$50.86 & 11.15$\pm$0.85\\
\enddata
\tablecomments{Basic properties of our stacked, X-ray detected, and combined bins.}
\end{deluxetable*}

\bibliography{References}{}
\bibliographystyle{aasjournal}

\end{document}